\documentclass[prb,preprint,showpacs,superscriptaddress,groupedaddress]{revtex4}  % for
\usepackage{makeidx}

\usepackage[colorlinks,hyperindex]{hyperref}
\usepackage[absolute]{textpos}
\usepackage{graphicx}  % needed for figures
\usepackage{epstopdf}
\usepackage{dcolumn}   % needed for some tables
\usepackage{bm}        % for math
\usepackage{amssymb}   % for math
\hypersetup
{
	colorlinks,%
	citecolor=blue,%
	linkcolor=blue,%
	urlcolor=blue,%
}

\makeindex

\begin{document}
	
	%\headno{PHYSICAL SCIENCES: Applied Physical Sciences}
	
	\begin{textblock}{14}(7,1)
		\noindent Published: \href{http://journals.aps.org/prb/abstract/10.1103/PhysRevB.92.235206}{Physical Review B 92, 235206 (2015)}
	\end{textblock}

\title{Coupling between Phonon-Phonon and Phonon-Impurity Scattering: A Critical Revisit of the Spectral Matthiessen's Rule}
%\input author_list.tex       % D0 authors (remove the first 3 lines
% of this file prior to submission, they
% contain a time stamp for the authorlist)
% (includes institutions and visitors)
\author{Tianli Feng, Bo Qiu, and Xiulin Ruan} %$\footnote{Email: ruan@purdue.edu}$}
\email{ruan@purdue.edu}
\affiliation{School of Mechanical Engineering and the Birck Nanotechnology Center, Purdue University, West Lafayette, Indiana 47907-2088, USA}

\date{\today}

\pacs{66.70.Df, 61.72.-y, 63.20.kp, 63.20.kg}

\begin{abstract}
	
	The spectral Matthiessen's rule is commonly used to calculate the total phonon scattering rate when multiple scattering mechanisms exist. Here we predict the spectral phonon relaxation time $\tau$ of defective bulk silicon using normal mode analysis based on molecular dynamics and show that the spectral Matthiessen's rule is not accurate due to the neglect of the coupling between anharmonic phonon-phonon scattering $\tau_a^{-1}$ and phonon-impurity scattering $\tau_i^{-1}$. As a result, the spectral Matthiessen's rule underestimates the total phonon scattering rate, and hence overestimates the thermal conductivity $\kappa$ of mass-doped and Ge-doped silicon by about 20-40\%. We have also directly estimated this coupling scattering rate, so-called coupled five-phonon scattering $\tau_{\rm couple}^{-1}$, and achieved good agreement between $\tau_a^{-1}+\tau_i^{-1}+\tau_{\rm couple}^{-1}$ and the total scattering rate $\tau_{tot}^{-1}$.
	
	% Furthermore, the importance of the impurity bonding to $\kappa$, which is often ignored in the Fermi's Golden Rule as well as in the Matthiessen's rule is investigated. We find that for 1\%-Ge doped Si, weakening and strengthening the Ge-Si bonding strength can produce an additional 50\% and 90\% reduction to $\kappa$, respectively. The weak Si-Ge bond makes Ge atoms ``rattle" in the Si matrix and results in an even lower $\kappa$ than vacancy doping.
	
\end{abstract}

\maketitle

The spectral Matthiessen's (M's) rule is a general rule to estimate the total spectral scattering rate when multiple scattering mechanisms exist at the same time\,\cite{Ziman_book}. In most solids, phonon transport is governed by phonon-phonon scattering $\tau_{a,\lambda}^{-1}$, phonon-impurity scattering $\tau_{i,\lambda}^{-1}$, phonon-boundary scattering $\tau_{b,\lambda}^{-1}$, etc., where $\lambda$ specifies a phonon mode $\lambda=(\mathbf{k},\nu)$ with wave vector $\mathbf{k}$ and dispersion branch $\nu$. By adding all of these factors, the spectral M's rule gives the total scattering rate as $\tau_{tot,\lambda}^{-1} = \tau_{a,\lambda}^{-1} +\tau_{i,\lambda}^{-1} +\tau_{b,\lambda}^{-1} +\cdots$. This scattering rate plays a crucial rule in predicting thermal conductivity $\kappa$ based on the Boltzmann transport equation (BTE) $\kappa_z=\frac{1}{V}\sum_\lambda v_{z,\lambda}^2 c_\lambda\tau_\lambda$, where $v_z$ is the projection of the phonon group velocity along the transport $z$ direction, $V$ is the volume, $c_\lambda$ is phonon specific heat per mode\,\cite{note_spheat}, and the summation is done over all the resolvable phonon modes\,\cite{Turney2009prb1}. This approach has been applied to predict the thermal conductivities of isotope-rich semiconductors\,\cite{Lindsay2013prb,Lindsay2008jp,Lindsay2012prl}, alloys\,\cite{Garg2011prl,He2012pccp,Tian2012prb}, nanostructures\,\cite{Mingo2003prb1,Martin2009prl}, etc.\,\cite{Feng2014Jn}. However, as an empirical rule, the M's rule assumes that the scattering mechanisms are independent, which was usually not verified in those calculations employing them. The spectral M's rule has not been examined yet, although the failures of the conventional gray M's Rules, i.e., $\sum\mu_j^{-1}=\mu_{tot}^{-1}$ or $\sum\sigma_j^{-1}=\sigma_{tot}^{-1}$ for electrical transport\,\cite{Ziman_book, Takeda1981EL} and $\sum\Lambda_j^{-1}=\Lambda_{tot}^{-1}$ or $\sum\kappa_j^{-1}=\kappa_{tot}^{-1}$ for phonon transport\,\cite{Turney2010jap,Luisier2013apl} have been studied extensively, where $\mu$, $\sigma$, and $\Lambda$ are the electron mobility, electrical conductivity and effective phonon mean free path, respectively. Take phonon transport for instance, the correct approach is to first obtain the spectral total relaxation time using the spectral M's rule $\tau_{tot,\lambda}^{-1}=\sum_{j}\tau_{j,\lambda}^{-1}$, and then derive the thermal conductivity using the Boltamann transport equation (BTE) $\kappa_{tot}\sim\int c(\omega)v_g(\omega)/v_p^2(\omega)\omega^2\tau_{tot,\lambda}(\omega)d\omega$, where $c$, $v_g$, and $v_p$ are the phonon specific heat, group velocity, and phase velocity, respectively. It can be conveniently shown that only if all the $\tau_j$'s have the same $\omega$ dependence, one can first use BTE to obtain the partial thermal conductivity due to one scattering mechanism $\kappa_j\sim\int c(\omega)v_g(\omega)/v_p^2(\omega)\omega^2 \tau_{j,\lambda}(\omega)d\omega$, and then use the gray M's rule to obtain the same total thermal conductivity $\kappa_{tot}^{-1}=\sum_j \kappa_j^{-1}$, i.e., the gray M's rule is valid. However, in general the $\tau_j$'s have different $\omega$ dependencies. Therefore, the failure of the gray M's rule can be expected, while the spectral M's rule has been assumed to be valid all the time.  Thus in this paper, our objectives are to (1) predict the spectral phonon scattering rate $\tau_{tot,\lambda}^{-1}$ and thermal conductivity $\kappa$ without touching the detailed phonon scattering processes or the spectral M's rule, and (2) examine the accuracy of the spectral M's rule and provide physical interpretation and quantitative correction.

%It is note that the gray M's rule $\kappa_{tot}^{-1}\sim\sum_j\left[\int c(\omega)v_g(\omega)/v_p^2(\omega)\omega^2\tau_j(\omega)d\omega\right]^{-1}$ can be derived from the spectral M's Rule $\kappa_{tot}\sim\int c(\omega)v_g(\omega)/v_p^2(\omega)\omega^2[\sum_j\tau_j^{-1}(\omega)]^{-1}d\omega$ only when all the $\tau_j$'s have the same $\omega$ dependence,  Therefore, the failure of the gray M's rule is within expectation under the framework of spectral M's rule which was assumed to be valid all the time.

We take defective bulk Si as an example and calculate $\tau_{a,\lambda}^{-1}$, $\tau_{i,\lambda}^{-1}$, and $\tau_{tot,\lambda}^{-1}$ in three independent ways. $\tau_{a,\lambda}^{-1}$ is obtained by performing the phonon normal mode analysis (NMA)\,\cite{Feng2014Jn,Thomas2010prb, Ladd1986prb,Koker2009prl, McGaughey2004prb, Henry2008jctn, Feng2015jap} on pristine silicon, in which only phonon-phonon scattering occurs. $\tau_{i,\lambda}^{-1}$ is calculated from the Tamura's formalism, and $\tau_{tot,\lambda}^{-1}$ is calculated by NMA on the defective silicon. The spectral M's rule is examined by comparing $\tau_{a,\lambda}^{-1}+\tau_{i,\lambda}^{-1}$ to $\tau_{tot,\lambda}^{-1}$. To examine the accuracy of these scattering rates, we compare the thermal conductivity $\kappa$ predicted from $\tau_{a,\lambda}^{-1}+\tau_{i,\lambda}^{-1}$ or $\tau_{tot,\lambda}^{-1}$ with that from the Green-Kubo method based on MD.
The scattering rates $\tau_{a,\lambda}^{-1}$ and $\tau_{tot,\lambda}^{-1}$ were obtained by performing the following NMA on pristine silicon and on defective silicon respectively,
\begin{eqnarray}
	q_\lambda(t) &=& \sum_\alpha^3\sum_b^n\sum_l^{N_c}\sqrt{\frac{m_b}{N_c}}u_\alpha^{l,b}(t)e_{b,\alpha}^{\lambda *}\exp\left[i\mathbf{k}\cdot \mathbf{r}_0^l\right], \label{normal-mode} \\
	\Phi_\lambda(\omega) &=& \left| \mathcal{F} [\dot{q}_\lambda(t)] \right|^2= \frac{C_\lambda}{(\omega-\omega_\lambda^A)^2+(\tau_{\lambda}^{-1})^2/4}. \label{SEDfunction} %(\omega_\lambda^{A\ 2}+\Gamma_\lambda^2)q_{\lambda,0}^2
\end{eqnarray}
Here, $u_\alpha^{l,b}(t)$ is the $\alpha$th component of the time dependent displacement of the $b$th basis atom in the $l$th unit cell, $e$ is the phonon eigenvector, $\mathbf{r}_0$ is the equilibrium position, $\mathcal{F}$ denotes the Fourier transformation, and $\Phi_\lambda(\omega)$ is called spectral energy density. $C_\lambda$ is a constant for a given $\lambda$. $\Phi_\lambda(\omega)$ is a Lorentzian function with peak position $\omega_\lambda^A$ and full width $\tau_\lambda^{-1}$ at half maximum. By MD simulation the time dependent atomic velocity $\dot{u}$ is obtained and substituted into Eqs.\,(\ref{normal-mode}) and (\ref{SEDfunction}) to obtain the spectral phonon scattering rate $\tau_\lambda^{-1}$. With this method, we can obtain $\tau_{a,\lambda}^{-1}$ of pristine silicon and $\tau_{tot,\lambda}^{-1}$ of impurity-doped silicon independently. In the evaluation of $\tau_{tot,\lambda}^{-1}$, the intrinsic lattice anharmonicity and the extrinsic impurity are treated as a combined perturbation to the phonon normal modes. This method does not touch the details of the scattering processes or the spectral M's rule.

From the second-order perturbation theory\,\cite{Klemens_book, Ziman_book}, Tamura gave the isotope scattering rate by fermi's golden rule (FGR)\,\cite{Tamura1983prb}
\begin{equation}
	\label{eq_isotope_tot}
	\tau_{i,\lambda}^{-1}= \frac{\pi}{2N_c}\omega_\lambda^2\sum_b^n  \sum_{\lambda'\neq\lambda} g_b |\mathbf{e}_b^\lambda\cdot \mathbf{e}_b^{\lambda'*}|^2 \delta(\omega_\lambda-\omega_{\lambda'}),
\end{equation}
where $g_b=\sum_\beta f_{\beta b} (1-m_{\beta b}/\bar{m}_b)^2 $ characterizes the magnitude of mass disorder, with $\beta$, $f_{\beta b}$, and $\bar{m}_b$ indicating the isotope types, the fraction of isotope in the $b$th basis atom, and the average atom mass at the $b$th basis. Equation \,(\ref{eq_isotope_tot}) is equivalent to $\pi g \omega_\lambda^2 D(\omega_\lambda)/2$ for cubic lattice structures, where $D(\omega)$ is the normalized density of states. The Tamura's formalism was first derived for the calculation of isotope scattering rate but recently has been applied to many other impurities with bonding change\,\cite{Garg2011prl,Tian2012prb,He2012pccp,Kundu2011prb,Li2012prb2}. In the last part of the paper we study the contribution of the impurity bonding strength to the total scattering rate. In the long wavelength approximation (LWA)\,\cite{Klemens_book, Kaviany_book, Tamura1983prb}, Eq.\,(\ref{eq_isotope_tot}) is reduced to the $\sim\omega^4$ relation, $\tau_i^{-1}=\frac{V_c n_c(\Delta m)^2}{4\pi v_g v_p^2m_c^2}\omega^4$, where $V_c$ is the volume of a unit cell, $n_c$ is the concentration of the impurities, $v_g$ and $v_p$ are the group and phase velocities of phonon, respectively.

\begin{figure}[tbph]% add gragh
	\centering
	\includegraphics[width= 4.5in]{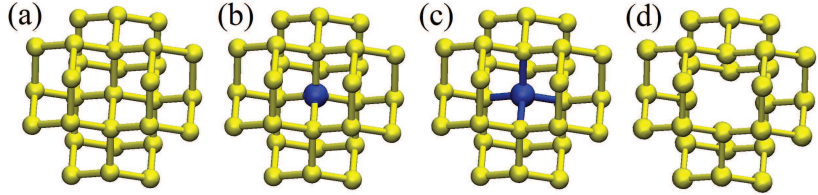}
	\caption{(Color online) Sketches of the lattice structures of (a) the pristine \emph{c}-Si, (b) the mass-doped \emph{c}-Si (${}^{m_i}$Si-Si), (c) the ${}^{73}$Ge-doped \emph{c}-Si (${}^{73}$Ge-Si), and (d) the vacancy-doped \emph{c}-Si (V-Si).}\label{fig_defect}
\end{figure}

We investigate the pristine \emph{c}-Si, the mass-doped \emph{c}-Si (${}^{m_i}$Si-Si), the ${}^{73}$Ge-doped \emph{c}-Si (${}^{73}$Ge-Si), and the vacancy-doped \emph{c}-Si (V-Si) bulks\,\cite{Shiga2014jjap, Wang2014msmse} at classical 300 K with the sketches of the lattice structures shown in Fig.\,\ref{fig_defect}. Here the ${}^{m_i}$Si-Si is to substitute some of the original Si atoms with mass $m_i$ while keeping the bonds unchanged. The NMA and the Tamura's formalism rely on MD simulations and lattice dynamics (LD) calculations, which are conducted in LAMMPS\,\cite{LAMMPS} and GULP\,\cite{GULP}, respectively. All the scattering rates are calculated based on the Tersoff interatomic potential\,\cite{Tersoff1989}. 
% Quantum correction\,\cite{TBMD_wang,Lee1991prb,Turney_QC} is required to match the quantities from NMA, at the classical temperature $T_{MD}$, with those from Tamura's Formulism at quantum temperature $T$. All $\kappa$ values reported hereafter are after quantum corrections, $\kappa=\kappa_{MD}dT_{MD}/dT$.
The domain size and total simulation time are set as $8\times8\times8$ conventional cubic cells and $10\ ns$ to eliminate the size and time effects\,[19]. Each time step is set as $\Delta t=0.5\ fs$ to resolve all the phonon modes. From the simulation results, it is found that one impurity affects at most the motions of its nearest (2.3 {\AA}) and second nearest (3.8 {\AA}) neighbors because of the approximate tight binding force in silicon. In our simulation, the impurities were randomly distributed with the distance between each of the two defects being larger than 11 {\AA} to ensure the defects do not influence each other. Three or more independent simulations are conducted for each case to minimize the statistical error. In the lattice dynamics calculation we employed a $\mathbf{k}$ grid of $96\times96\times96$ to obtain results as accurately as possible, since Eq.\,(\ref{normal-mode}) requires the evaluation of delta functions.

\begin{figure}[tbph]% add gragh
	\centering
	\includegraphics[width= 4.5in]{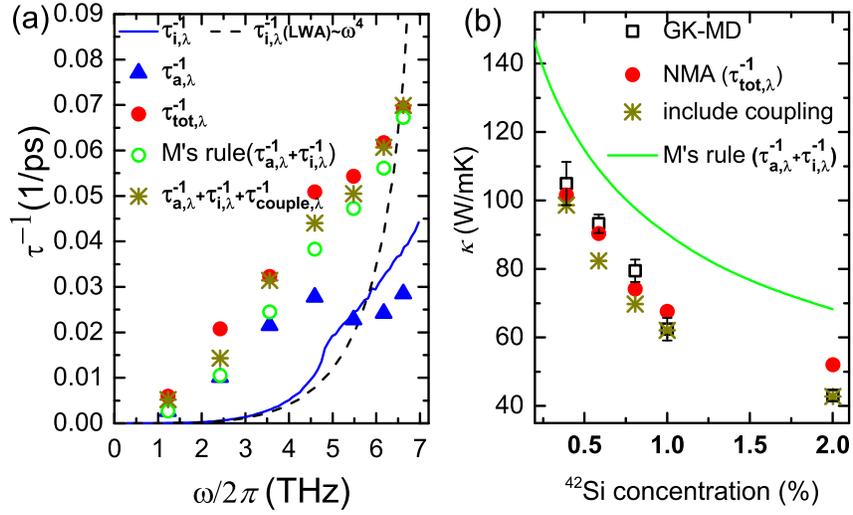}
	\caption{(Color online) (a) The phonon scattering rates of the TA mode in the [100] direction in pristine \emph{c}-Si and 0.4\% ${}^{42}$Si doped silicon. (b) The thermal conductivity of ${}^{42}$Si-Si calculated from four different methods as a function of the ${}^{42}$Si concentration.}\label{fig_relax}
\end{figure}

Figure \,\ref{fig_relax} (a) gives the phonon scattering rates of the TA mode in the [100] direction for 0.4\% ${}^{42}$Si-Si at classical 300 K. The phonon-phonon scattering rate $\tau_{a,\lambda}^{-1}$ scales as $\sim\omega^2$ at low frequencies, but deviates at higher frequencies. Such a trend was also seen in previous studies\,\cite{Henry2008jctn, Esf2011prb}. The $\tau_{i,\lambda}^{-1}$ calculated from Eq.\,(\ref{eq_isotope_tot}) is found to be exactly the same with $\pi g \omega_\lambda^2 D(\omega_\lambda)/2$ as mentioned above. At low frequency, $\tau_i^{-1}$ is about one order of magnitude lower than $\tau_a^{-1}$. As the frequency increases, $\tau_i^{-1}$ becomes even higher than $\tau_a^{-1}$ due to the increase of the density of states. We also note that the impurity scattering rate obeys the $\sim\omega^4$ relation given by the LWA [dashed black line] for the phonon frequency below $1.5$ THz. For higher frequency where phonon wavelength is short and comparable with the defect size, the Rayleigh scattering model breaks down giving way to the Mie scattering model and thus the $\sim\omega^{4}$ frequency dependence gradually fades with increasing frequency\,\cite{Klemens1955bond}. Interestingly, we find that the spectral M's rule, $\tau_a^{-1}+\tau_i^{-1}$ [open green circle], underestimates the phonon total scattering rates [solid red circle] by 10\%-50\% for different frequencies. To ensure this discrepancy is not due to the different domain sizes used in the NMA and Tamura's formula, we have performed both the MD simulations and FGR calculations in the domain of 16$\times$16$\times$16 unit cells and the same-size $\mathbf{k}$ mesh, respectively. We have found that such discrepancy indeed exists, especially in the mid-frequency range. Actually the impurity scattering rates do not vary much when the $\mathbf{k}$-mesh size changes from 96$\times$96$\times$96 to 16$\times$16$\times$16 since the latter is already dense enough to obtain a good phonon density of states. In our work, all the calculations are done based on the same Tersoff potential, and the comparison between the results of different methods is self-consistent. Thus, the conclusions still hold although the dispersion has discrepancy with experiments.

\begin{figure}[h]% add gragh
	\centering
	\includegraphics[width= 4in]{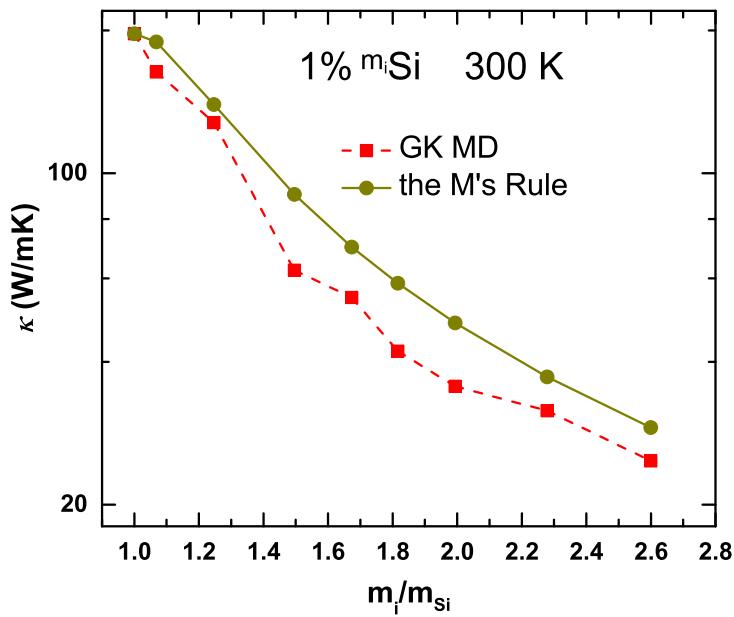}
	\caption{(Color online) The comparison of the thermal conductivity calculated from GK-MD and the M's rule for 1\% ${}^{m_i}$Si-Si at classical 300 K.}\label{massratio}
\end{figure}

The thermal conductivity $\kappa$ as a function of the ${}^{42}$Si concentration at classical 300 K is shown in Fig.\,\ref{fig_relax}(b). The NMA and GK are both based on classical equilibrium molecular dynamics and the same interatomic potential.
%Also, the same quantum correction is done for both of them. The difference is that the GK method uses the fluctuation-dissipation theorem and is considered an exact method when the system is classical, while the NMA depends on perturbation theory and thus only works for slightly perturbed systems.  Hence, the NMA can be benchmarked against GK. 
In our calculation results, their agreement is good (within 5\%). As seen in Fig.\,\ref{fig_relax}(b) for the mass doped bulk silicon, the NMA thermal conductivity values (red circles) match excellently with GK values (black squares). For pristine \emph{c}-Si, our Green-Kubo and NMA methods give consistent thermal conductivity values.
% $\kappa$ of $168.98\pm18.56$ W/mK and $161$ W/mK after quantum corrections\,\cite{note_QC}, respectively, which agree reasonably with the experimental value $156$ W/mK\,\cite{experimentSi}. 
In contrast, the $\kappa$ calculated from the spectral M's rule [green line] has about 20\%-40\% overestimation. This overestimation has also been observed in the doped silicon with a broad range of mass (28-73) at a concentration of 1\%, as seen in Fig.\,\ref{massratio}.

\begin{figure}[tbph]% add gragh
	\centering
	\includegraphics[width= 4in]{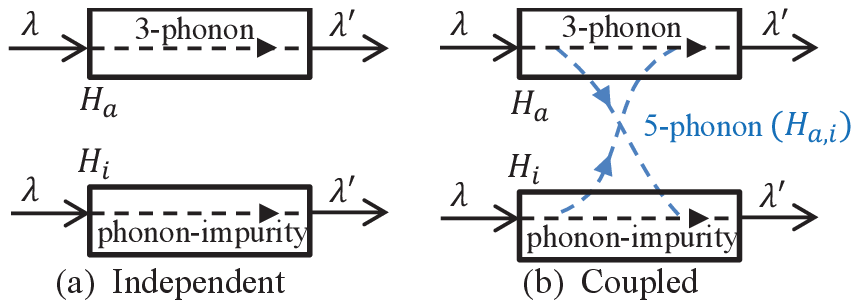}
	\includegraphics[width= 4in]{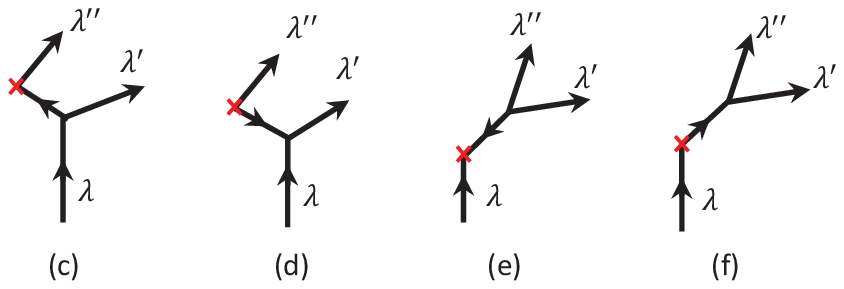}
	\caption{(Color Online) Brief sketches for illustrating (a) the independent scattering mechanisms and (b) the coupled scattering mechanisms. The sketches of coupling scattering are (c), (d), (e) and (f) with detailed description found in Refs.\,\cite{Carruthers1962pr1,Carruthers1961rmp}. The $\times$ represents the scattering by impurity. }\label{Ms_fig}
\end{figure}

The physical mechanism for the inaccuracy of the spectral M's rule is explored from the second-order perturbation theory\,\cite{Carruthers1962pr1}. The phonon scattering operator and rate for a defective material are described as
\begin{eqnarray}
	H_{\rm scatt} \! = \! H_a \!+\! H_i \!+\! (H_a\!+\!H_i)(E\!+\!H_0\!+\!i\varepsilon)^{-1}(H_a\!+\!H_i), \label{eq_operator} \\
	\lefteqn{\tau_{tot,\lambda}^{-1} = \tau_{a,\lambda}^{-1}+\tau_{i,\lambda}^{-1}+\tau_{{\rm couple},\lambda}^{-1},} \hspace{3.1in} \label{eq_new_rule}
\end{eqnarray}
where $H_0$ is the harmonic lattice Hamiltonian, and $\varepsilon$ is a positive infinitesimal\,\cite{Carruthers1962pr1}. The first two terms $H_a$ and $H_i$ are the perturbation Hamiltonians from intrinsic anharmonicity and extrinsic impurity, leading to intrinsic anharmonic phonon-phonon scattering $\tau_{a,\lambda}^{-1}$ and extrinsic phonon-impurity scattering $\tau_{i,\lambda}^{-1}$, respectively. The former includes the intrinsic three-phonon, four-phonon, five-phonon processes, etc., and the latter involves two phonons. The third operator in Eq.\,(\ref{eq_operator}), which was usually ignored by researchers, represents the coupling between $H_a$ and $H_i$ and may involve five or more phonons. To the lowest order of the coupling, the coupled five-phonon scattering (three phonons in the three-phonon scattering and the two phonons in the impurity scattering) provides additional channels for one mode to scatter to the other mode and thus increases the scattering rate, as shown in Figs.\,\ref{Ms_fig}(a) and (b). The detailed sketches for the coupled five-phonon process are shown in Figs.\,\ref{Ms_fig}(c) to (f). Note that this coupled five-phonon process is different from the intrinsic five-phonon process which has already been included in the term $\tau_{a,\lambda}^{-1}$. The term ``coupling" is used because the transitions occur between the intermediate quantum states of three-phonon process and impurity-phonon process with detailed sketches shown in Refs.\,\cite{Carruthers1962pr1,Carruthers1961rmp}. This coupling is calculated by the second-order perturbation theory and is different from the meaning of ``interplay" discussed in Refs.\,\cite{Lindsay2013prb2,Lindsay2012prb} where the spectral M's rule was still used\,\cite{note_coupling}.

To roughly estimate the contribution of the coupled five-phonon scattering, we applied the approximate expression derived by Carruthers from fermi's golden rule\,\cite{Carruthers1962pr1},
\begin{equation}
	\label{eq_new_rule_max}
	\tau_{{\rm couple},\lambda}^{-1}=\frac{3g}{4}\left(3\left(\frac{\omega}{\Delta \omega}\right)^2+\frac{3}{4}+\frac{6\pi\bar{\omega}^4}{\tau_{tot,\lambda}^{-1}\omega_0^3}\right)\tau_{a,\lambda}^{-1},
\end{equation}
where $\Delta\omega$ measures ``the `lack' of energy conservation by the intermediate phonons"\,\cite{Carruthers1962pr1} in the five-phonon process, and $\omega_0$ is Debye frequency\,\cite{explanation}. This coupled five-phonon scattering rate, however, has, to our knowledge, never been evaluated. We substitute the $\tau_{tot}^{-1}$ obtained from NMA into Eq.\,(\ref{eq_new_rule_max}) to estimate the coupling scattering $\tau_{{\rm couple},\lambda}^{-1}$ and check the agreement between $\tau_{tot,\lambda}^{-1}$ and $\tau_{a,\lambda}^{-1}+\tau_{i,\lambda}^{-1}+\tau_{{\rm couple},\lambda}^{-1}$. By including the estimated coupling scattering rate $\tau_{\lambda,{\rm couple}}^{-1}$, a good agreement is achieved for both $\tau$ and $\kappa$ as seen in Fig.\,\ref{fig_relax}. The frequency dependence of $\tau_{{\rm couple},\lambda}^{-1}$ in Fig.\,\ref{fig_relax}(a) is explained by Eq.\,(\ref{eq_new_rule_max}) as follows. At low frequencies (1-5 THz), $\tau_{{\rm couple},\lambda}^{-1}$ increases with frequency due to the increasing $\tau_{a,\lambda}^{-1}$, while at higher frequencies (6-7 THz), it decreases with frequency since the increasing $\tau_{tot,\lambda}^{-1}$ brings down the third term in the bracket. Physically, at higher frequencies the large density of states allows phonon states to transit into other states quickly by impurity scattering, and thus the intermediate states required in the coupling scattering are probably difficult to produce. Generally, the maximum $\tau_{a,\lambda}^{-1}$ occurs at the mid-frequencies where the phonons have relatively high $\tau_{a,\lambda}^{-1}$ as well as low density of states.

%Thus, the maximum coupling effect in general occurs in the frequency range in which $\tau_{a,\lambda}^{-1}$) is strong while $\tau_{i,\lambda}^{-1}$ is as weak as possible. According to this rule, our result in Fig.\,\ref{fig_relax} can be explained: the maximum coupling effect at the mid-frequencies (2-5 THz) where the phonons have the highest $\tau_{a,\lambda}^{-1}$ and $\tau_{a,\lambda}^{-1}/\tau_{i,\lambda}^{-1}$.

\begin{figure}[tbph]% add gragh
	\centering
	\includegraphics[width= 4in]{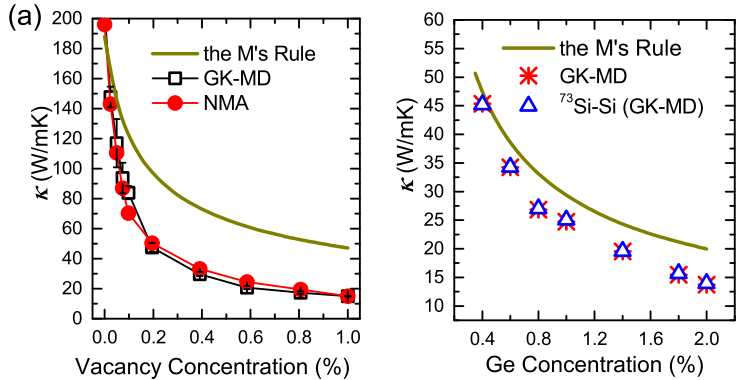}
	\caption{(Color online) Thermal conductivity of (a) V-Si and (b) ${}^{73}$Ge-Si calculated from GK-MD, NMA, and the spectral M's rule as a function of vacancy or Ge concentration at classical 300 K. The blue triangles label $\kappa$'s of ${}^{73}$Si-Si as references.}\label{VSi}
\end{figure}

Figure \,\ref{VSi} (a) shows $\kappa$ of V-Si as a function of vacancy concentration calculated from different methods. We find that even for bond-missing defects our NMA [red circle] presents excellent agreement with GK-MD [black square], indicating that treating the impurities and anharmonicity as one combined perturbation to calculate total scattering rates is reasonable. The spectral M's rule [yellow line] over-predicts $\kappa$ of V-Si by about 100-150\% at a vacancy concentration of 0.2-1\%. The discrepancy comes from two aspects: (1) the spectral M's rule neglects the coupling between phonon-phonon and phonon-defect scattering rates as elaborated earlier, and (2) the Tamura's formalism only captures the scattering by mass disorder while ignoring the bond changes. As for ${}^{73}$Ge-Si shown in Fig.\,\ref{VSi} (b), the spectral M's rule over predicts about 20\%-40\% of $\kappa$. Most of this discrepancy comes from the coupling since we find that ${}^{73}$Ge-Si and ${}^{73}$Si-Si have almost the same thermal conductivity, indicating that the Ge-Si bond provides negligible scattering compared to the mass disorder introduced by Ge atoms. In addition, the over prediction of $\kappa$ by the spectral M's rule is also seen in the high Ge concentration range\,\cite{Carruthers1962pr1,Abeles1962pr}.

For two materials having the same light-doping level ($\tau_{i,\lambda}^{-1}\ll\tau_{a,\lambda}^{-1}$), the coupling strength, defined by $\tau_{couple,\lambda}^{-1}/\tau_{tot,\lambda}^{-1}$ is approximately $1/\tau_{a,\lambda}^{-1}$ and is higher for the higher-$\kappa$ material which has a lower $\tau_{a,\lambda}^{-1}$. On the other hand, if the doping level is high ($\tau_{i,\lambda}^{-1}\gg\tau_{a,\lambda}^{-1}$), the coupling strength $\sim \tau_{a,\lambda}^{-1}/\tau_{i,\lambda}^{-1}$ is higher for the lower-$\kappa$ material which has a higher $\tau_{a,\lambda}^{-1}$.

\begin{figure}[tbph]% add gragh
	\centering
	\includegraphics[width= 4in]{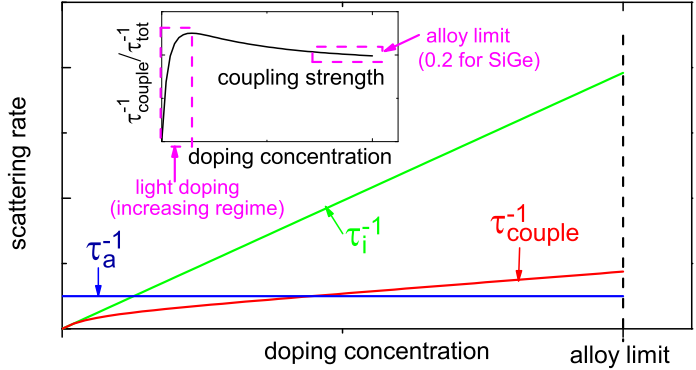}
	\caption{(Color online) A sketch to show the scattering rates and the coupling effect (inset) as a function of doping level.}\label{general}
\end{figure}

For general materials, the phonon scattering rates as a function of doping concentration are shown in Fig.\,\ref{general}. The coupling scattering rate initially increases rapidly with doping in the light doping regime and then increases linearly and more slowly in the heavy doping regime. As a result, a maximum of the coupling strength occurs when the system transits from the light to heavy doping. For example, for silicon doped with Ge in our work as shown in Fig.\,\ref{VSi}(b), the coupling strength increases to 0.4 as the doping level increases to 2\%. On the other hand, at the concentration of 50\% which is in the alloy limit, the coupling strength is about $\tau_{couple}^{-1}/\tau_{tot}^{-1}\approx 4\tau_{a}^{-1}/\tau_{tot}^{-1}\approx 4(1/156)/(1/7)\approx 0.2$. Here we used the approximation of $\tau_{couple}^{-1}\approx 10g\tau_{a}^{-1}$\,\cite{Carruthers1962pr1} with $g\approx0.4$ in SiGe alloy and the fact that pristine silicon and SiGe alloy have the thermal conductivities of 156 W/mK and 7 W/mK, respectively. Carruthers \textit{et al.} hypothesized that this coupled five-phonon scattering caused the over estimation of $\kappa$ for SiGe alloy in early years\,\cite{Carruthers1962pr1,Abeles1962pr}, though a quantitative evaluation was not done in their work. The concept of the coupling effect can be extended to all doped material systems, and the general trends should be similar to Fig.\,\ref{general}. For example, the coupling strength in PbTe/Se alloy is estimated to be about 9\%, which may account for the overestimation in Ref.\,\cite{Tian2012prb}. In Ref.\,\cite{Garg2011prl}, Garg \emph{et al.} included the coupling implicitly by calculating the three-phonon scattering rates in a large SiGe alloy supercell using fully-quantum density functional perturbation theory. Although the five-phonon processes were implicitly included in prior calculations of the total phonon scattering rates, we have isolated the scattering rate due to five-phonon processes only.

To conclude, without touching the details of the phonon scattering processes, we have used the NMA approach to predict the thermal properties of defective materials more accurately than the spectral M's rule. The spectral M's rule is found to over-predict the phonon relaxation time and thermal conductivity because the spectral M's rule does not take into account the coupling between anharmonic phonon-phonon scattering and impurity scattering. Our results demonstrate one system which has strong coupling between different scattering mechanisms and estimate the coupling scattering rates with good quantitative accuracy. Such coupling exists in many different systems of solids, and can be explored for lower $\kappa$ as well as higher $ZT$ for thermoelectrics.

%To further study how much the impurity bonding strength influences the thermal transport, we calculate $\kappa$ of 1\% ${}^{73}$Ge-Si while artificially changing the Ge-Si bonding strength from weak to strong. On the stronger bond side, doubling the bond strength gives about 20\% reduction to $\kappa$. Towards the infinite strength limit, the impurity bonds affect the whole lattice and completely break the original harmonicity, leading to infinitely low thermal conductivity\,\cite{suppl_bonding}. On the weaker bond side, half strength produces about 40\% reduction to $\kappa$, and the reduction saturates to around 50\% when the bond strength is weakened to a certain extent. At the weak bond limit, the Ge atom behaves like a ``rattler" in the cave surrounded by the four Si neighbors. We note that this 1\% ``rattler" Ge atoms give a thermal conductivity even 20\% lower than that of 1\% vacancy doping. This may provide a guidance to figure out the validity of the controversial ``rattling'' model for thermal conductivity reduction in filled skutterudites\,\cite{baoling, rattling1994}.

We would like to thank Ajit Vallabhaneni and Frank Brown for proofreading the manuscript, and Yan Wang for helpful discussions. Simulations were performed at the Purdue Network for Computational Nanotechnology (NCN). The work was supported by the National Science Foundation (Award No. 1150948).

\bibliography{../../../bibfile}

\begin{thebibliography}{40}
\bibitem[{not({\natexlab{a}})}]{note_spheat}
\bibinfo{note}{To consistent with the GK method based classical MD simulations,
  in this work the value of $c_\lambda$ is taken as $k_B$ under the classical
  Boltzmann distribution.}
\end{thebibliography}

\end{document}